\documentclass[twocolumn]{openjournal}
\usepackage{booktabs}
\usepackage{xcolor}
\usepackage{hyperref}
\begin{document}
\newcommand{\grantcomment}[1]{{\color{blue}\bf[GR:  {#1}]}}

\title{Spectroscopic and X-ray Modeling of the \\Strong Lensing Galaxy Cluster MACS J0138.0-2155}

\author{Abigail S. Flowers$^{1,2}$, Jackson H.~O'Donnell$^{1,2}$, Tesla E.~Jeltema$^{1,2}$, Vernon Wetzell$^{3}$,\\ and M.~Grant Roberts$^{1,2}$}

\affiliation{$^1$ University of California, Santa Cruz, Santa Cruz, CA 95064, USA}
\affiliation{$^2$ Santa Cruz Institute for Particle Physics, Santa Cruz, CA 95064, USA}
\affiliation{$^3$ Department of Physics \& Astronomy
University of Pennsylvania
Philadelphia, PA 19104, USA}

\begin{abstract}
		\noindent We model the total mass and galactic substructure in the strong lensing galaxy cluster MACS J0138.0-2155 using a combination of Chandra X-ray data,  Multi-Unit Spectroscopic Explorer (MUSE) spectroscopy, and Hubble Space Telescope imaging.  MACS J0138.0-2155 lenses a source galaxy at $z=1.95$ which hosts two strongly lensed supernovae, Requiem and Encore.  We find MACS J0138.0-2155 to have an X-ray temperature of $6.7\pm0.4$ keV and a velocity dispersion of cluster member galaxies of $718^{+132}_{-182}$ km s$^{-1}$. These lead to the mass estimates for the cluster of $M_{500} = 5.2^{+1.5}_{-1.2} \times 10^{14} M_\odot$ from the X-ray results and $M_{200} \approx 3.6^{+2.0}_{-2.7} \times 10 ^{14} M_{\odot}$ from the velocity dispersion results.  The round morphology of the X-ray emission indicates that this cluster is relaxed with an ellipticity within the lensing region of $e=0.12\pm0.03$.  Using 18 of the brightest, non-blended, quiescent galaxies, we fit the cluster specific Faber-Jackson relation, including a set of 81 variations in the analysis choices to estimate the systematic uncertainties in our results. We find a slope of $\alpha = 0.26 \pm 0.06 (\mathrm{stat.}) \pm 0.03 (\mathrm{sys.})$ with an intrinsic scatter of $\Delta \sigma = 31^{+8}_{-6} (\mathrm{stat.}) \pm 4 (\mathrm{sys.})$ km s$^{-1}$ at a reference velocity dispersion of $\sim 220$ km s$^{-1}$.  We also report on significant galaxies along the line-of-sight potentially impacting the lens modeling, including a massive galaxy with stellar velocity dispersion of $236 \pm 3$ km s$^{-1}$ which lies close in projection to the central cluster galaxy.  This galaxy is part of a small group at a slightly higher redshift than the cluster.
\end{abstract}

\section{Introduction}

Strong lensing by galaxy clusters enable wide ranging science, including studies of the highest-redshift galaxies, lensed supernova, independent measurements of the Hubble constant $H_0$ and the dark matter equation of state parameter $w$, detection of dark matter substructure, and constraints on the nature of dark matter.  However, an accurate cluster mass modeling must account for cluster substructure and baryonic components.  In this work, we focus on the remarkable strong lensing cluster MACS J0138.0-2155 (hereafter MACS0138) and provide spectroscopic analysis of cluster member galaxies and other galaxies in the field as well as X-ray analysis of the intra-cluster medium (ICM).

MACS0138 is a particularly special strong lensing cluster at $z=0.337$, with a highly-magnified, lensed source galaxy at $z=1.95$ \citep{Newman2018}.  
The source galaxy has now exhibited not just one but two strongly-lensed supernova, Requiem and Encore, both Type 1a \citep{Rodney2021, Pierel2024}.  The presence of a radial arc and an overall relaxed mass distribution also make MACS0138 and ideal candidate for strong lens mass modeling.

In this paper we present the analysis of Multi Unit Spectroscopic Explorer (MUSE) integral-field spectroscopy and Chandra X-ray data of MACS0138.  Among other results, we present 
kinematic modeling of cluster member galaxies and derive a cluster specific Faber-Jackson relation \citep{FJ76} between the 
member galaxy stellar velocity dispersions and their luminosities.  This relation allows for modeling of the member galaxy contributions to the lensing signal and will be incorporated in the lens modeling presented in a forthcoming paper (O'Donnell et al.~in preparation).
We also model the velocity dispersion of galaxies within the cluster as a whole and the X-ray temperature, both benchmarks for the total cluster mass. 

\section{Data and Data Reduction}

\subsection{MUSE Spectroscopy}
Much of the data used for this analysis was taken with the European Southern Observatory's Multi-Unit Spectroscopic Explorer (MUSE) \citep{Bacon:2010}, 
installed on UT4 of the Very Large Telescope (VLT).
This integral field spectrograph provides 3-dimensional data cubes, which can be resolved as ordinary 2-dimensional images, for which each pixel contains full spectroscopic information. While this analysis largely does not make use of this spatial information beyond taking individual sources from it, this data does allow us the freedom of choosing after the fact the region from which we would like to collect our sample.

MACS0138 was observed with MUSE for 2.9 ksec on September 6, 2019 as part of program 0103.A-0777. The observations were taken in wide field mode, providing a FOV of approximately 1'x1' with a ``spaxel'' scale of 0.2", and a spectral coverage of roughly 475-935 nm. This work relies on a fully reduced data cube of these observations obtained from the ESO data archive, reduced with the standard \texttt{muse} pipeline version 2.8 \citep{musepipe}.

\subsection{HST Imaging} \label{sec:hst_imaging}
\begin{figure}
    \centering
    \includegraphics[width=1\linewidth]{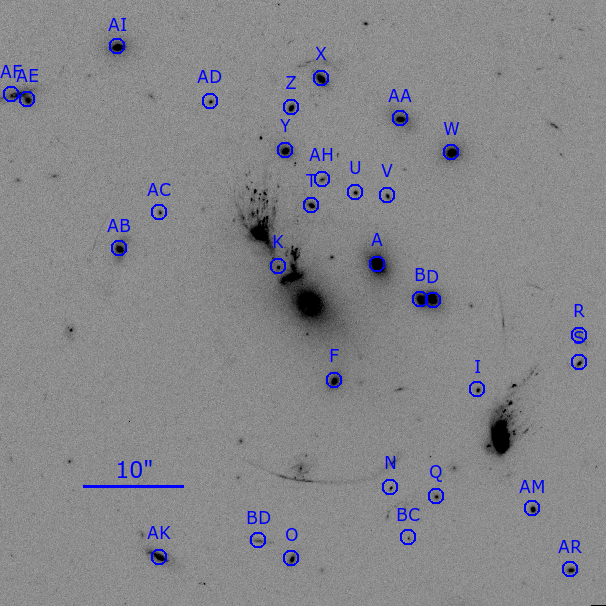}
    \caption{HST F555W image with with galaxies in our spectroscopic sample listed in Table \ref{tab:alldata} labeled. The circular galaxy regions have radius 0.8" and the image is roughly 1' per side.}
    \label{fig:hst}
\end{figure}

Hubble imaging on MACS0138 was used to model the light profiles and shapes of the cluster galaxies.  In this work, these were used to determine the member galaxy luminosities.  We use the HST ACS/WFC imaging taken in the F555W band (Proposal ID 14496). The data were taken from the Mikulski Archive for Space Telescopes (MAST), specifically the calibration level 3 image which was processed with DrizzlePac 3.6.2.  Figure \ref{fig:hst} shows the central region of the cluster in the HST F555W image with galaxies in our final spectroscopic catalog labeled.  We choose the F555W band as it is the closest to the spectral range we use; all of the galaxies used in our Faber-Jackson relation fit are quiescent and do not show spectral emission lines.

Galaxy photometry was modeled using the \texttt{tractor} algorithm \citep{tractor}. In brief, the ``black point'' (sky background) and error are measured from several different blank patches across the field, which all give consistent values.
The point spread function (PSF) is measured from a nearby star which is modeled in \texttt{tractor} as a sum of three Gaussians. Galaxies in the field were modeled as follows: If they have no other galaxies within 1" of their center, a 1"x1" cutout around the galaxy is taken. If there are multiple galaxies within that distance, a larger cutout including all the nearby galaxies is taken, with at least 0.5" of room on each side. In this case the light profile from each galaxy is fit jointly. Most galaxies are modeled as a sum of an exponential and de Vaucouleur profile \citep{deVau53}. For small and faint galaxies, we use an exponential only; of the galaxies in our final sample, this applied to galaxies `Q' and `R'.  Fluxes are converted to magnitudes using the AB mag zero-point computed from the PHOTFLAM and PHOTPLAM keywords in the reduced image and extinction corrected taking the galactic extinction values from the NASA/IPAC Extragalactic Database (NED) \footnote{https://ned.ipac.caltech.edu/Documents/References/\\ExtinctionCalculators}.

\subsection{Chandra X-ray Data}
MACS0138 was observed by Chandra in June 2015 with the ACIS-I instrument in VFAINT mode for 24.73 ksecs (ObsID 17186).  These data were reduced using the Mass Analysis Tool for Chandra (MATCha) \citep{Hollowood19} running CIAO 4.13 \citep{ciao} as part of the X-ray follow up of clusters found in the Dark Energy Survey \citep{Kelly24}.  MATCha automates the data reduction and spectral analysis of clusters of galaxies observed by Chandra including reprocessing the data from the level 1 event files, removing particle background flares, creating images and exposure maps, and detecting and removing point sources.  MATCha then measures the centroid position and extracts and fits the X-ray spectrum within several apertures to derive the X-ray temperatures and luminosities.  The detailed procedures can be found in \cite{Hollowood19} and \cite{Kelly24}.

In this paper, we consider the temperatures and luminosities measured by MATCha and use the exposure-corrected image to fit the surface brightness profile and ellipticity as discussed in Section \ref{sec:xray_results}.  The X-ray spectra are fit with XSPEC \citep{xspec} to an diffuse hot gas emission model including elemental line emission and the affects of line of sight absorption; specifically we use the wabs*mekal model with the abundance set to $0.3Z_{\sun}$ and the galactic neutral hydrogen column density in the direction of the cluster taken from the \textit{HEASOFT} tool \textit{nH} which takes the weighted average of the densities in \cite{Kalberla} and \cite{Dickey}.  X-ray temperatures and luminosities are fit in apertures of radius 500 kpc, $r_{2500}$, $r_{500}$, and core-cropped $r_{500}$ with a core size of $0.15 r_{500}$.  Here $r_\Delta$ is taken to be the radius within which the average density is $\Delta$ times the critical density, and this radius is estimated iteratively based on the X-ray temperature and the scaling of radius and temperature from \cite{arnaud05}.

\section{Methodology}

\subsection{Spectroscopic modeling} \label{sec:specmodeling}

Spectroscopic analysis of each object was accomplished in two steps. First,
a preliminary redshift estimate of each object was obtained with the online tool \texttt{MARZ} \citep{marz}. Second, more detailed spectral fitting was performed with the standard package \texttt{pPXF} \citep{Cappellari:2004,Cappellari:2017,Cappellari:2023} to measure both a more precise redshift and the stellar velocity dispersion of each galaxy.

For each galaxy photometrically fit by \texttt{tractor} as described in Sec.~\ref{sec:hst_imaging}, a 1D spectrum was extracted by combining all `spaxels' within the elliptical effective radius. These labeled spectra were combined and fed into \texttt{MARZ}; where they were visually inspected, assigned redshifts, and assigned quality flags from 0 to 4. This work utilizes objects with quality flags of 3 and 4. We do not use the central galaxy in the analysis in this paper, and we also do not include four, blue jellyfish galaxies (galaxies undergoing ram-pressure stripping with prominent tails); these galaxies show star formation and signs of active stripping affecting their dynamics and will be discussed in a future work.  

\begin{figure}
    \centering
    \includegraphics[width=1\linewidth]{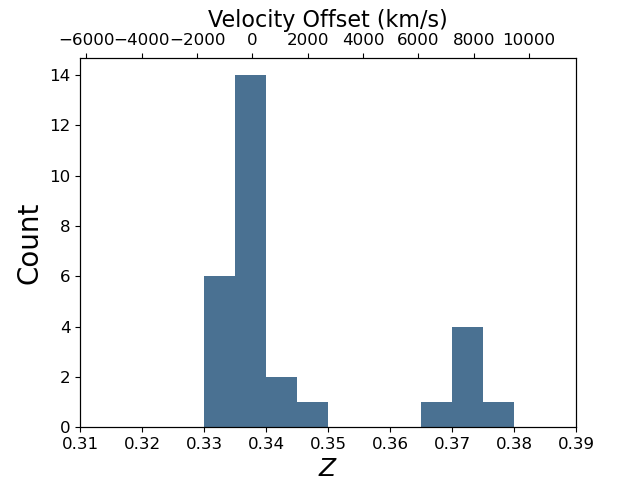}
    \caption{Histogram of confident redshifts of galaxies in the MUSE field as fit by \texttt{MARZ} with redshifts near the cluster redshift.}
    \label{fig:marz_histogram}
\end{figure}

These cuts yielded a sample of 30 galaxies with good redshifts listed in Table \ref{tab:alldata}. 
 Of these, 23 are clearly at the redshift of MACS0138 and a set of 5 galaxies are at slightly higher redshift. A histogram of the redshift distribution is shown in Figure \ref{fig:marz_histogram}. An additional two galaxies (N and R) are at significantly higher redshifts than the cluster, but are included here as they lie near the lensing region and are foreground objects to the lensed source galaxy.

More detailed spectral modeling was done for the bright, non-blended cluster members using \texttt{pPXF}.  This analysis used the available spectra in the rest-frame wavelength range 3850-4500 \r{A}, spanning the Ca H\&K and G-band absorption features.
\,As input stellar templates to \texttt{pPXF}, we use data release 3 of the X-Shooter Library (XSL)~\citep{Verro:2022}, which contains spectra of 630 stars. In order to prevent over-fitting with low-quality templates, we exclude any XSL spectrum with a mean SNR per pixel lower than 3 in the wavelength range used for spectral fitting. When fitting a spectrum from the \texttt{MUSE} data, \texttt{pPXF} creates a linear combination of template stellar spectra to best match the input galaxy's spectrum. We first fit a bright member galaxy's spectrum, and use the template combination provided by \texttt{pPXF} for that galaxy as a fixed ``optimal template'' for the rest of the cluster members in our sample.
Since the spectral resolution of X-Shooter exceeds that of MUSE, these templates were convolved with a Gaussian line spread function to match the intrinsic resolution of our observed spectra.

To ensure that our measured velocity dispersions are robust to systematic effects introduced by the choice of fit parameters, we repeat these \texttt{pPXF} fits for each galaxy with 81 separate configurations of \texttt{pPXF} parameters. This approach is similar to that taken in \citet{Shajib:2023}, and is described in detail in Section~\ref{sec:FJmethod}.
Redshifts and velocity dispersions measured by this method are reported in Table \ref{tab:alldata}. 
As part of the more detailed spectral analysis, we also attempted to fit the spectra of likely cluster members whose \texttt{MARZ} quality was 2, which resulted in adding one galaxy `AC' for which we were able to get a good \texttt{pPFX} fit. We also include in the table the velocity dispersion of galaxy `A', discussed in Section \ref{sec:los}, which is a massive galaxy at a slightly higher redshift important to the lens modeling.

\subsection{Faber-Jackson Relation}
\label{sec:FJmethod}
As described above, we use the MUSE spectroscopy to determine the peculiar velocities and velocity dispersions of individual cluster member galaxies.  We use the former to determine the velocity dispersion of galaxies within the cluster and the latter along with the HST photometry to fit the scaling relation between galaxy luminosity and stellar velocity dispersion, the Faber-Jackson relation. 

Firstly, in mapping the Faber-Jackson relation of the cluster, we must make a selection for which member galaxies to include. We select cluster galaxies, with sufficient spectral signal-to-noise to reliably measure a stellar velocity dispersion via absorption features. Ultimately, the 18 galaxies we chose to include in this relation were those which were brighter than $m_{\mathrm{F555W}} = 23.5$, within 3,000 km s$^{-1}$ of the central recessional velocity of the cluster, which did not appear to be spatially blended in the available MUSE data, and had a pPFX fit $\chi^2$ less than 2.5  and an error on the stellar velocity dispersion less than 75 km s$^{-1}$.  We do not include the central galaxy in the Faber-Jackson relation; cluster central galaxies have a different formation history than satellite galaxies and do not follow the same Faber-Jackson relation \cite[e.g.][]{vonderlinden2007}.  We separately model the dynamics of the central galaxy and jointly fit these with the lens modeling in a forthcoming paper (O’Donnell et al. in preparation).

For each of the 18 galaxies in our selection, we extract the spectroscopic data from within a 0.8 arcsec radius around the central location of the galaxy in the MUSE observation and take that as this galaxy's spectrum. Then, using the method described above, we use the \texttt{pPXF} library to obtain each galaxy's peculiar velocity and stellar velocity dispersion.

In fitting a galaxy, \texttt{pPXF} requires a spectral template, which it can either build based on matching the target galaxy's spectrum by combining multiple spectral templates (here using stellar spectra from the XSL), or by taking an existing template as an input.

\begin{figure}
    \centering
    \includegraphics[width=1\linewidth]{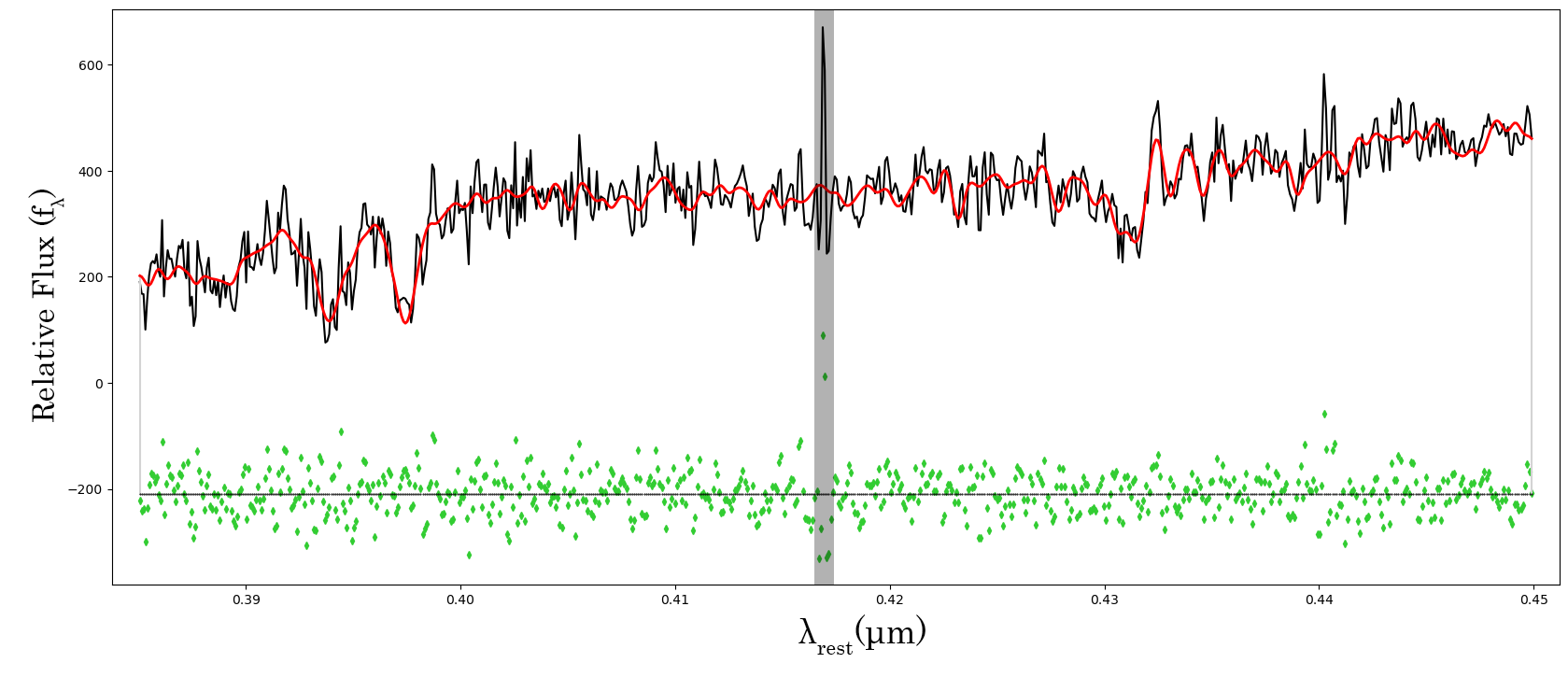}
    \caption{MUSE spectrum of the template galaxy used in the nominal Faber-Jackson fit, galaxy W (black), overlaid with \texttt{pPXF}'s best fitting model (red) and residuals (green). The spectrum is shown from 3850\r{A} to 4500\r{A} and includes visible Calcium H \& K and G-band absorption features. The feature shown in the gray region is an artifact of the sky subtraction.}
    \label{fig:ppxffit}
\end{figure}

To keep our nominal run self-consistent, we make a choice of template galaxy from our sample and build a template based on it that is then used for all other galaxies in the sample. We choose our template galaxy as one that is particularly bright and well-defined. Specifically, we chose galaxy W for this purpose; the spectrum for galaxy W is show in Figure~\ref{fig:ppxffit}. \texttt{pPXF} also requires a degree of polynomial used to fit the spectrum, which for our nominal run was set to 4. The systematics associated with both of these choices were explored as described below.

Once we have the magnitude and velocity dispersion for each galaxy in our sample, we fit the Faber-Jackson relation following a similar methodology as \cite{Bergamini:2019}.  We assume a power law relation between galaxy stellar velocity dispersion and luminosity  using:

\begin{equation}
    \sigma = \sigma_{\mathrm{ref}}  \left( \frac{L}{L_0} \right) ^\alpha
    \label{eq:Faber-Jackson}
\end{equation}

where $\sigma$ is a given galaxy's velocity dispersion, $\sigma_{\mathrm{ref}}$ is a constant for the reference velocity dispersion at a specified magnitude (here taken to be $m_{F555W} = 20$), $L$ is the luminosity of the member galaxy, $L_0$ is the reference luminosity (corresponding to a magnitude of $m_{F555W} = 20$), and $\alpha$ is the power-law slope. 

We fit the Faber-Jackson relation following the methodology outlined in Appendix B of \cite{Bergamini:2019}. Specifically, we employ a Markov Chain Monte Carlo (MCMC) method with \texttt{emcee}'s affine invariant ensemble sampler \citep{GoodmanWeare,emcee} to find the values of $\sigma_{\mathrm{ref}}$ and $\alpha$ which best fit our population. Additionally, we fit the variable $\Delta\sigma$ representing the intrinsic scatter in the relation. These three variables were fit using a log prior, constrained to the ranges $\alpha  \in (0, 0.6)$ km s$^{-1}$, $\sigma_{ref} \in (50, 500)$, and $\Delta\sigma \in (0, 200)$ km s$^{-1}$. Our log-likelihood function follows Appendix B of~\citet{Bergamini:2019}. 

In addition to this nominal fit, we perform a systematics procedure to test the dependence of our findings on the particular choices made in the analysis. Here, we repeat the spectral modeling for our entire sample of 18 galaxies and re-fit the Faber-Jackson relation, each time iterating over a range of choices made in the initial analysis. We iterate over a total of 4 variables, each with 3 different options, giving a total of 81 different runs. 

The first variable iterated over was the degree of polynomial used by \texttt{pPXF} to fit the spectrum, set to degree 4 in our nominal run, and either 3, 4, or 5 in our systematics. The second variable was the radius of the spectral extraction aperture, set as either 0.5", 0.8", or 1". The third variable was the choice of which galaxy spectrum should be used to form the template from the combination of XSL spectral templates, iterating between the 3 brightest, well-behaved galaxies (here referred to as W, AB, and AI). The final variable iterated over was the portion of the XSL used to make the template spectrum, iterating between one half of the library (randomly selected), the complimentary other half, and the full library. 

Each of the 81 possible permutations of these variables was run through \texttt{pPXF} with the population of 18 galaxies as described above to derive a new sample of stellar velocity dispersions, and the resulting spread provides us with the systematic error for our measurements.  Fits with $\chi^2$ greater than 2.5  or an error on the stellar velocity dispersion greater than 75 km s$^{-1}$ were removed from the sample for that run. We then re-fit the Faber-Jackson relation for each of the 81 realizations giving a range of values for our power-law fit parameters.  In Table \ref{tab:alldata}, we list the central 68\% range on the velocity dispersions of the individual galaxies from these systematics runs.

\subsection{Cluster Velocity Dispersion}
\label{sec:sigmethod}

Alongside the Faber-Jackson analysis, we also present an estimate for the velocity dispersion of the cluster based on peculiar velocities of member galaxies. To do this, we take a slightly larger population of 23 galaxies whose peculiar velocities within the cluster are well-measured using \texttt{MARZ} (see Section~\ref{sec:specmodeling}). Many of these galaxies are the same as those used in the Faber-Jackson modeling. In general, the \texttt{MARZ} determined redshifts are very consistent with those measured by pPFX with some small deviations (see Table~\ref{tab:alldata}). 
As our \texttt{MARZ} dataset includes more redshift values than our limited \texttt{pPXF} run, \texttt{MARZ} redshifts were used for the cluster velocity dispersion.

We fit the cluster velocity dispersion following the methodology outlined in \textcite{Wetzell2019} and originally detailed in \citet{Beers1990}. Namely, we use the biweight location statistic to estimate the central redshift of the cluster 
as well as the biweight scale and gapper methods for determining the cluster's velocity dispersion. These statistics have been shown to be accurate for small sample sizes and robust to non-Gaussian distributions and outliers \citep{Beers1990}. 
We estimated the cluster biweight location twice: once with the full selection of galaxies and once with a smaller selection of only galaxies whose velocity offsets were within $\pm$3,000 km s$^{-1}$ of the cluster's central redshift based on the initial cluster biweight location, which reduces the population to 23. 

We then use this final sample of 23 member galaxies to estimate the velocity dispersion of galaxies within the cluster. 
We employ a bootstrapping method applied to both the biweight location and gapper statistics to estimate the uncertainties in the velocity dispersion. Specifically, we resample the data with replacement 1,000 times and recalculate the velocity dispersion with each resampling. 
We take the median of the bootstrap trials as our velocity dispersion estimate and the 68\% width as the uncertainty, and we compare this with the velocity dispersion for the nominal sample.

\section{Results}

\subsection{Faber-Jackson Fit}

\begin{figure}
    \centering
    \includegraphics[width=1\linewidth]{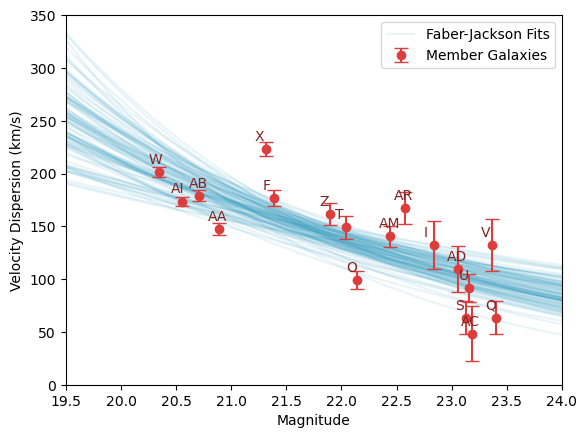}
    \caption{HST F555W magnitude versus velocity dispersion for our selection of cluster member galaxies. Also shown are the derived Faber-Jackson relations showing the final 100 results from the MCMC chain.}
    \label{fig:Faber-Jackson}
\end{figure}

\begin{figure}
    \centering
    \includegraphics[width=1\linewidth]{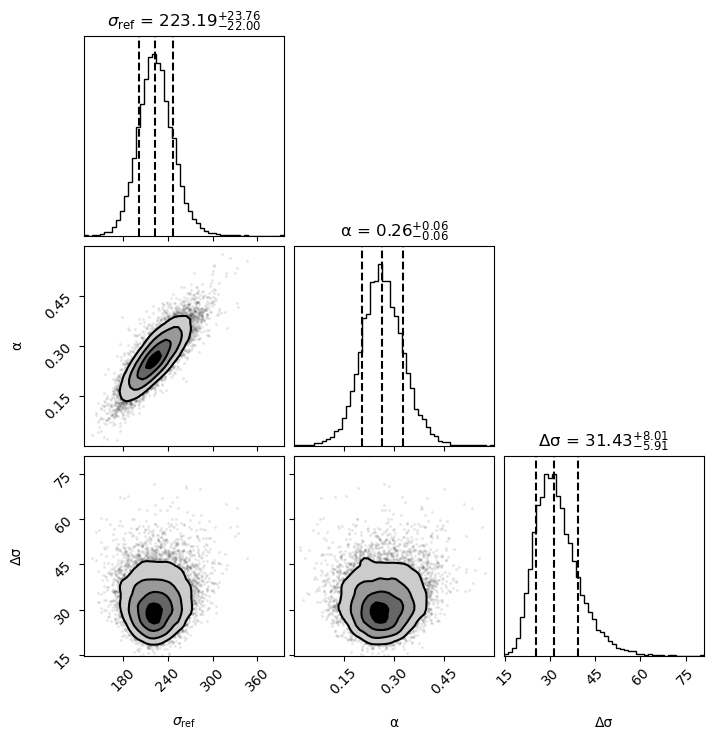}
    \caption{Corner plot showing the results of the MCMC fitting of the Faber-Jackson relation for the variables $\sigma_{\mathrm{ref}}$ (reference dispersion for m = 20), $\alpha$ (power-law slope), and $\Delta\sigma$ (intrinsic scatter); the histograms show the median values and the 68\% confidence interval.}
    \label{fig:Corner}
\end{figure}

Our results for the Faber-Jackson relation in MACS0138 are shown in Figure~\ref{fig:Faber-Jackson} with fit values and uncertainties listed in Table \ref{tab:Results}.
Both the nominal best-fit and the median of the MCMC chains give a slope of $0.26$; taking the central 68\% of the parameter distribution around the median as the $1 \sigma$ confidence interval (e.g. the 16th and 84th percentiles), we find $\alpha = 0.26 \pm 0.06$.  For the intrinsic scatter we find median value and uncertainties of $\Delta \sigma = 31^{+8}_{-6}$ km s$^{-1}$ and a reference velocity dispersion at $m=20$ of $\sigma_{ref}=223^{+24}_{-22}$ km s$^{-1}$. In all cases, the median parameter values are very consistent with the best-fit values.  Figure \ref{fig:Corner} shows corner plots of the parameter distributions for our nominal fits.  These show some degeneracy between the slope and reference velocity dispersion and a small, non-Gaussian tail toward higher intrinsic scatter, but in general are well behaved.

\begin{table*}[]
    \centering
    \begin{tabular}{c||c|c|c|c|c|c|c}
         Galaxy & RA & DEC & F555w & MARZ & pPXF & Velocity & Systematics\\
         Label & & & Magnitude & Redshift & Redshift & Dispersion (km s$^{-1}$) & Deviation (km s$^{-1}$) \\
    \midrule
        A$\dag$* & 24.5137 & -21.9244 & 19.24 & 0.3706 & 0.3707 & 236$\pm$3 & [236, 238] \\
        B$\dag$ & 24.5124 & -21.9254 & 20.84 & 0.3330 & -- & -- & --\\
        D$\dag$ & 24.5120 & -21.9254 & 20.75 & 0.3379 & -- & -- & --\\
        F & 24.5150 & -21.9276 & 21.39 & 0.3314 & 0.3331 & 177$\pm$7 & [160, 177]\\
        I & 24.5107 & -21.9279 & 22.84 & 0.3355 & 0.3361 & 133$\pm$23 & [109, 188] \\
        K$\dag$ & 24.5166 & -21.9245 & 19.20 & 0.3377 & -- & -- & -- \\
        N$\dag$* & 24.5133 & -21.9306 & 23.71 & 0.4583 & -- & -- & -- \\
        O & 24.5162 & -21.9326 & 22.14 & 0.3351 & 0.3357 & 99$\pm$8 & [96, 111] \\
        Q & 24.5119 & -21.9309 & 23.40 & 0.3347 & 0.3355 & 63$\pm$16 & [42, 73] \\
        R$\dag$* & 24.5076 & -21.9264 & 23.24 & 0.8301 &  -- & -- & --\\
        S & 24.5076 & -21.9271 & 23.13 & 0.3367 & 0.3370 & 64$\pm$16 & [33, 100] \\
        T & 24.5156 & -21.9228 & 22.04 & 0.3371 & 0.3372 & 149$\pm$11 & [142, 164] \\
        U & 24.5143 & -21.9224 & 23.16 & 0.3337 & 0.3348 & 92$\pm$13 & [90, 114] \\
        V & 24.5134 & -21.9225 & 23.37 & 0.3355 & 0.3363 & 132$\pm$25 & [125, 173] \\
        W & 24.5114 & -21.9213 & 20.35 & 0.3336 & 0.3346 & 201$\pm$5 & [201, 208] \\
        X & 24.5154 & -21.9192 & 21.31 & 0.3381 & 0.3382 & 223$\pm$7 & [208, 232] \\
        Y$\dag$ & 24.5164 & -21.9212 & 20.50 & 0.3530 & -- & -- & -- \\
        Z & 24.5162 & -21.9201 & 21.90 & 0.3398 & 0.3392 & 162$\pm$10 & [159, 187] \\
        AA & 24.5130 & -21.9204 & 20.89 & 0.3327 & 0.3339 & 147$\pm$6 & [135, 149] \\
        AB & 24.5214 & -21.9240 & 20.71 & 0.3377 & 0.3377 & 179$\pm$5 & [162, 174] \\
        AC* & 24.5202 & -21.9230 & 23.19 & 0.3365 & 0.3367 & 48$\pm$26 & [72, 142] \\
        AD & 24.5187 & -21.9199 & 23.06 & 0.3425 & 0.3411 & 110$\pm$22 & [41, 83] \\
        AE$\dag$ & 24.5241 & -21.9198 & 21.63 & 0.3376 & -- & -- & -- \\
        AF$\dag$* & 24.5246 & -21.9197 & 21.60 & 0.3752 & -- & -- & -- \\
        AH$\dag$ & 24.5153 & -21.9220 & 23.05 & 0.3453 & -- & -- & -- \\
        AI & 24.5214 & -21.9184 & 20.55 & 0.3384 & 0.3382 & 173$\pm$5 & [159, 179] \\
        AK$\dag$* & 24.5202 & -21.9325 & 22.36 & 0.3715 & -- & -- & -- \\
        AM & 24.5090 & -21.9312 & 22.44 & 0.3391 & 0.3387 & 141$\pm$10 & [127, 159] \\
        AR & 24.5079 & -21.9329 & 22.58 & 0.3380 & 0.3379 & 167$\pm$15 & [130, 169] \\
        BC$\dag$* & 24.5127 & -21.9320 & 24.51 & 0.3715 & -- & -- & -- \\
        BD$\dag$* & 24.5172 & -21.9321 & 23.22 & 0.3695 & -- & -- & -- \\
    \end{tabular}
    \caption{Catalog of galaxies with confident spectroscopic redshift measurements. Column 1 lists the galaxy labels as shown in Figure~\ref{fig:hst}. 
 Columns 2-4 list the positions and magnitudes in the HST F555W imaging. Column 5 gives the \texttt{MARZ} determined redshifts; while Columns 6 and 7 list the pPFX redshifts and stellar velocity dispersions for the 18 galaxies used modeling the Faber-Jackson relation plus the massive interloping galaxy A. Column 8 gives the central 68\% range of the velocity dispersion for each galaxy from our systematics tests described in Section \ref{sec:FJmethod}. $\dag$ indicates galaxies not used in the Faber-Jackson modeling, and $*$ indicates galaxies not used in finding the cluster velocity dispersion.}
    \label{tab:alldata}
\end{table*}

Alongside these nominal results we performed the systematics tests described in Section \ref{sec:FJmethod} to test the affects of our analysis choices on the resulting fit. 
The systematics analysis results in a total of 81 different sets of stellar velocity dispersion measurements for our sample of cluster galaxies.  We fit each of these 81 data sets using the same procedure as out nominal sample.  Table~\ref{tab:Results} lists the median and 68\% spread of the best-fit parameter values from the systematics runs.  The resulting values are very consistent within the errors with our nominal fit, though the median best-fit slope from the systematics run is slightly shallower.  The systematic uncertainties are found in all cases to be less than the statistical ones, but do add to the total uncertainty.  Taken together, we find the slope of the Faber-Jackson relation to be $\alpha = 0.26 \pm 0.06 
 (\mathrm{stat.}) \pm 0.03 
 (\mathrm{sys.})$.  The distributions of the slope, reference velocity dispersion, and intrinsic scatter in the systematics trials are shown in the Appendix.

\begin{table}
    \centering
    \begin{tabular}{c || c | c | c}
    Variable & Nominal & Nominal & Median of\\
     & Best-Fit & Median & Systematics Best-Fits \\
    \midrule
    $\sigma_{\mathrm{ref}}$ (km s$^{-1}$) & 221 & $223^{+24}_{-22}$ & $213^{+7}_{-7}$ \\
    & & &\\
    $\Delta\sigma$ (km s$^{-1}$) & 27 & $31^{+8}_{-6}$ & $29^{+4}_{-4}$\\
    & & &\\
    $\alpha$ & 0.26 & $0.26^{+0.06}_{-0.06}$ & $0.23^{+0.03}_{-0.03}$ \\
    \end{tabular}
    \caption{Table showing the results of the Faber-Jackson relation modeling.  Column 2 lists the best-fit parameters for the nominal run.  Column 3 gives the median of the MCMC chains along with the 16th and 84th percentile limits. Column 4 lists the median of the best fits for the systematics runs along with the spread in the medians.}
    \label{tab:Results}
\end{table}

\subsection{Cluster Velocity Dispersion}

Figure \ref{fig:marz_histogram} shows the distribution of galaxy redshifts for the galaxies in Table \ref{tab:alldata} using the \texttt{MARZ} redshifts.  As can be seen in this figure, the cluster appears as a main clump of 23 galaxies with a central redshift of 0.3367 from the biweight location statistic.  A secondary clump of 5 galaxies lies at a slightly higher redshift around 0.37; this clump includes the luminous galaxy A which lies close in projection to the center of the cluster (see Figure \ref{fig:hst}).

\begin{figure}
    \centering
    \includegraphics[width=1\linewidth]{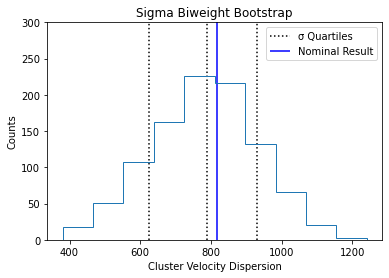}
    \includegraphics[width=1\linewidth]{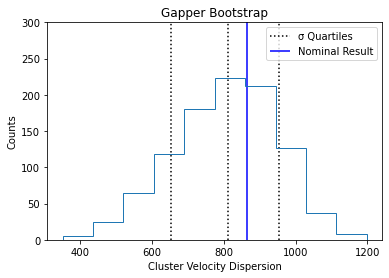}
    \caption{Histogram of bootstrap results for the biweight (top) and gapper (bottom) methods of determining the cluster velocity dispersion, including vertical lines indicating the nominal result, median of the bootstrap results, and the 68\% confidence interval.}
    \label{fig:sigma_bootstrap}
\end{figure}

Using the 23 galaxies with peculiar velocities within 3,000 km s$^{-1}$ of the cluster redshift, we estimate the cluster velocity dispersion using both biweight scale and gapper statistics. 
As described in Section \ref{sec:sigmethod} we employ a bootstrap procedure to estimate the uncertainties in the velocity dispersion, specifically using 1,000 random resamplings of the data with replacement. For the biweight method, we find a median velocity dispersion and 68\% confidence limits of $671^{+152}_{-166}$ km s$^{-1}$, and for the gapper method $718^{+132}_{-182}$ km s$^{-1}$. These are consistent with each other and with the nominal results with baseline sample. The spread of results found from bootstrapping for both methods are shown in Figure~\ref{fig:sigma_bootstrap}; in both cases these distributions are fairly Gaussian in shape.

\begin{table}[]
    \centering
    \begin{tabular}{c|| c | c}
     Dispersion & Nominal & Bootstrap \\
     Method & Result (km s$^{-1}$)& Median (km s$^{-1}$)\\
     \midrule
    Biweight & 706 & $671^{+152}_{-166}$\\
    & & \\
    Gapper & 761 & $718^{+132}_{-182}$ \\
    & & \\
    \end{tabular}
    \caption{Results for cluster velocity dispersion from the biweight and gapper methods. Column 2 gives the results for the nominal galaxy sample, while column 3 gives the median of the bootstrap trials along with the 16th and 84th percentile limits.}
    \label{tab:Cluster_Dispersion}
\end{table}

Using our estimate for the galaxy velocity dispersion of the cluster, we can, given some assumptions, estimate the $M_{200}$ mass of the cluster. The method we use, described in \citet{Evrard:2008} and used in \citet{Buckley-Geer:2011}, recognizes that in using solely the cluster's line-of-sight galactic velocity dispersion, we lack significant information about cluster shape, galactic orbits, and velocity biases. The relation is given by

\begin{equation}
    b_v^{1/\alpha}M_{200} = 10^{15}M_\odot \frac{1}{h(z)}\left(\frac{\sigma_{gal}}{\sigma_{15}}\right)^{1/\alpha}
    \label{eq:M200}
\end{equation}

where $b_v$ represents the unknown biases in the galactic velocities relative to the dark matter, $h(z)$ is the reduced Hubble constant at redshift $z$, $\sigma_{gal}$ is the estimated galactic velocity dispersion of the cluster, $\sigma_{15}$ is the reference velocity dispersion for a cluster of mass $M_{200} = 10^{15}  M_{\odot}$, and $\alpha$ is a dimensionless constant. Here, we adopt $\sigma_{15} = 1082.9 \pm 4$ km s$^{-1}$ and $\alpha = 0.3361 \pm 0.0026$ following~\citet{Buckley-Geer:2011} and~\citet{Evrard:2008}. For $h(z)$, we assume a flat $\Lambda$CDM cosmology with $h(z) = h_0\sqrt{\Omega_M (1 + z)^3 + \Omega_\Lambda}$ and take $\Omega_M =0.3$, $\Omega_\Lambda = 0.7$, and $h_0=0.69$.

Using Equation~\ref{eq:M200} and the gapper estimate for the velocity dispersion, we find $b_\nu^{1/\alpha} M_{200} = 3.6^{+2.0}_{-2.7} \times 10^{14} M_\odot$, where the errors include the uncertainties in both the velocity dispersion and in the parameters of Equation~\ref{eq:M200}.

\subsection{Significant Galaxies along the Line-of-Sight}
\label{sec:los}

\begin{figure}
    \centering
    \includegraphics[width=1\linewidth]{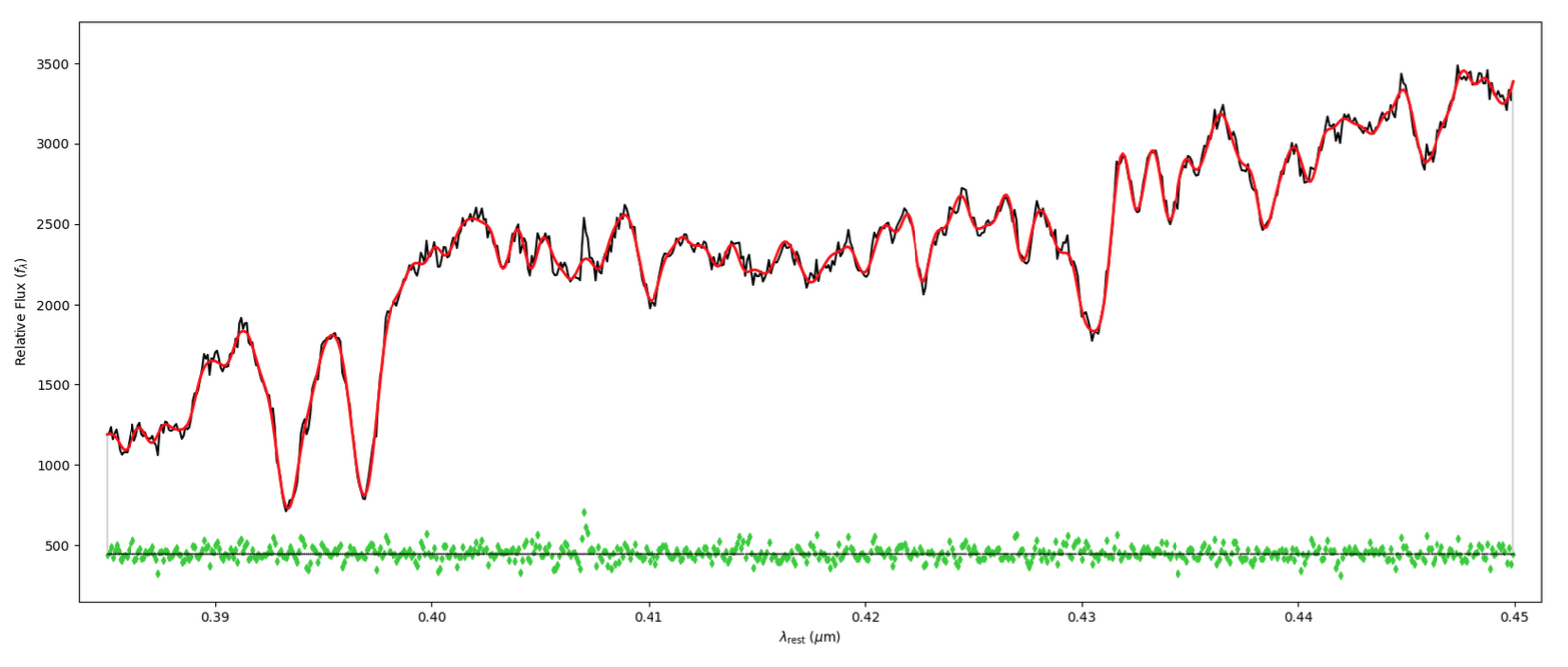}
    \caption{\texttt{MUSE} spectrum of galaxy A (black) overlaid with \texttt{pPXF}'s best fitting model (red) and residuals (green). The spectrum is shown from 3850\r{A} to 4500\r{A} and includes visible Calcium H \& K and G-band absorption features.}
    \label{fig:Aspec}
\end{figure}

As previously noted, some of the galaxies in our full redshift sample are not cluster members, but still lie along the line of sight to the lensed source and thus affect the lens modeling. 
Notably, galaxy A is the brightest in the field besides the central galaxy and one of the closest galaxies in projection to the central.
 Based on the \texttt{pPXF} analysis, this galaxy has a velocity dispersion of 236 km s$^{-1}$, corroborating that the galaxy is of high mass. The spectrum and pPFX fit for this galaxy are shown in Figure \ref{fig:Aspec}.  Along with A, we find four, fainter galaxies with similar redshifts, AF, AK, BC, and BD.

 Two additional galaxies in our spectroscopic sample N at $z=0.46$ and R at $z=0.83$ lie at higher redshifts, but are close in projection to lensed arcs.  

\subsection{X-ray Properties}
\label{sec:xray_results}

From the Chandra analysis, we find that MACS0138 has an $r_{500}$ X-ray temperature of $6.7\pm0.4$ keV and a soft band (0.5-2 keV) luminosity of $5.5\pm0.1 \times 10^{44}$ ergs s$^{-1}$, with an $r_{500}$ radius of $1089^{+41}_{-38}$ kpc.  The temperature is consistent for different aperture choices; for example, the $r_{2500}$ temperature is found to be $6.5^{+0.4}_{-0.3}$ keV at an $r_{2500}$ radius of $473^{+15}_{-13}$ kpc, and the core-cropped $r_{500}$ temperature is found to be $6.6^{+0.9}_{-0.8}$ keV.
Based on the $r_{500}$ temperature and using the $M-T$ relation from \citet{Mantz16}, we estimate the mass of the cluster to be $M_{500} = 5.2^{+1.5}_{-1.2} \times 10^{14} M_{\odot}$ where the uncertainties include both the X-ray temperature uncertainties and the uncertainties on the fit parameters in \citet{Mantz16}.

\begin{figure}
\includegraphics[width = 1\linewidth]{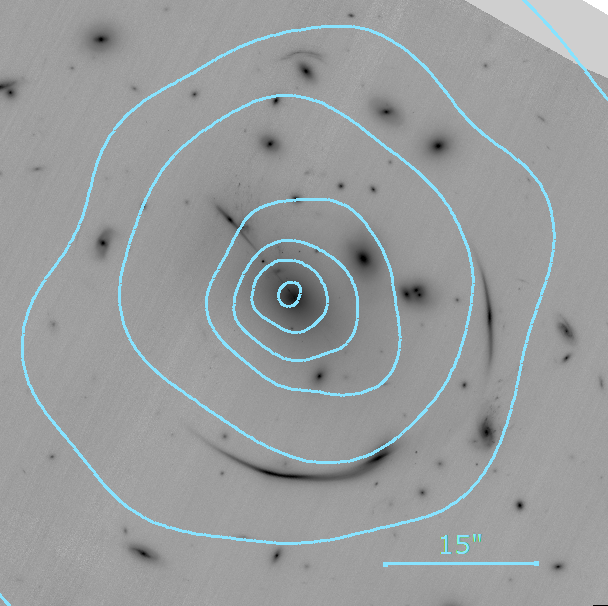}
\caption{Contours of X-ray emission overlaid on the JWST F200W image of MACS0138.  X-ray contours are log spaced and based on the Chandra 0.5-2 keV image which has been adaptively smoothed.}
\label{fig:xray}
\end{figure}

As can be seen in Figure \ref{fig:xray}, the morphology of the X-ray emission is round and relaxed. We fit the 2-dimensional X-ray surface brightness profile within the lensing region, an aperture with $r=19"$ the approximate radius of the observed giant arcs, using Sherpa \citep{sherpa}.  Specifically, we model the surface brightness as a 2D elliptical beta-model \citep{beta} of the form

\[
\Sigma = \Sigma_0 \left(1 + \frac{r}{r_c}\right)^{-3\beta+1/2}
\]
where $r_c$ is the core radius, $\Sigma_0$ is the central surface brightness, and $r$ is the appropriate elliptical radius
\[
r^2 = \frac{(x-x_0)^2(1-e)^2 + (y-y_0)^2}{(1-e)^2}
\]
with ellipticity $e$.  We find the ellipticity within the lensing region to be nearly circular with $e=0.12\pm0.03$.  We also find a small core radius $r_c = 10\pm1$ kpc, which is less than 1\% of $r_{500}$ for this cluster.  The small core radius could be indicative of a cool core cluster \citep[e.g.][]{Hudson10}.  In fact the MUSE spectra shows emission lines indicative of both star formation and AGN activity (O'Donnell et al.~in preparation).  Within the lensing region, we find $\beta = 0.43 \pm 0.01$.

\section{Discussion}

Both the X-ray temperature and cluster velocity dispersion indicate that MACS0138 is a massive cluster.  The mass estimate based on the X-ray temperature is somewhat higher at $M_{500} = 5.2^{+1.5}_{-1.2} \times 10^{14} M_{\odot}$ compared the the velocity dispersion estimate of $M_{200} = 3.6^{+2.0}_{-2.6} \times 10^{14} M_\odot$, but the two are consistent.  We note that our X-ray mass estimate is for a smaller radius ($r_{500}$) than the velocity dispersion mass ($r_{200}$).  The depth of the X-ray data is not sufficient to probe the temperature at larger radii, but given that MACS0138 appears to be isothermal within the radii probed, we would expect the $M_{200}$ mass to be about 40\% larger \citep{arnaud05, Umetsu20}, which is still within 2$\sigma$ of the velocity dispersion estimated mass. Using the lens model in the supplementary material of \cite{Rodney2021}, we estimate a cluster mass of $2.2 \times 10^{14} M_{\odot}$ with another $0.7 \times 10^{14} M_{\odot}$ in their BCG component. While on the low side, taking both of these components together this lensing mass is consistent with what we find.

\cite{Rodney2021} also find the cluster dark matter halo to be very elliptical with their lens model giving $e = 0.81^{+0.02}_{-0.13}$, which is at odds with the very round X-ray distribution for which we find $e=0.12\pm0.03$.  While the dark matter distribution is expected to be more flattended than the X-rays, which follow the isopotential surfaces, the difference is within a factor of 2-3 across several different mass distributions and 3D shapes \citep{Mcdaniel21}.

For the Faber-Jackson relation, \citet{Bergamini:2019} present results for three strong lensing clusters at similar redshifts to MACS0138 and find Faber-Jackson slopes of $\alpha = 0.28 \pm 0.02$, $0.27 \pm 0.03$, and $0.27 \pm 0.04$. Their results are very consistent with our result of $\alpha = 0.26 \pm 0.06 
 (\mathrm{stat.}) \pm 0.03 
 (\mathrm{sys.})$, though we note that their reference velocity dispersions as slightly higher than ours ($\sim 300$ km s$^{-1}$) and they include the BCG in their fits.  Our measured intrinsic scatter is also consistent with the clusters in \citet{Bergamini:2019} with the exception of Abell S1063 for which they measure a higher scatter.

 In the final stages of preparation of this paper, the preprint \citet{Granata24} was posted, which also fits the Faber-Jackson relation in MACS0138. Despite some differences in the two analyses, our results are completely consistent; they find $\alpha = 0.25 \pm 0.05$ and scatter $\Delta \sigma = 25^{+4}_{-6}$ km s$^{-1}$ at a similar reference velocity dispersion (206 km s$^{-1}$ compared to our 223 km s$^{-1}$). Furthermore, when fitting velocity dispersions of the same member galaxies, their analysis finds results largely consistent with those presented here. Their best fit velocity dispersions are all similar to ours and within the systematic and statistical uncertainties for every galaxy in common except AR and F for which we get larger velocity dispersion.  For F, considering the uncertainties our value is within $\sim 2 \sigma$ of what \citet{Granata24} find; for AR our velocity dispersion places it relatively high compared to the Faber-Jackson relation while the \citet{Granata24} value places it somewhat low. Differences in the two analyses include that we use a somewhat larger sample of galaxies (18 vs. 13) reflecting different quality cuts, they use a deeper MUSE cube obtained by their team, and they use F160W magnitudes compared to our use of F555W.  While they do investigate the affects of signal-to-noise on their velocity dispersion measurements, they do not employ the type of systematics test we do; our results indicate that their uncertainties may be underestimated.  However, the overall consistency of the results of these independent works speaks to their robustness.

After the submission of this work, the preprint \citet{Acebron:2025} was also posted using the same deeper MUSE data as \citet{Granata24}, and similarly conducts kinematic and X-ray analysis of cluster MACS0138. Their results for the Faber-Jackson slope excluding the BCG of $\alpha = 0.26\pm0.06$ is in good agreement with the findings presented here. Additionally, their findings for the X-ray temperature of 6-7 keV outside of 25 kpc, and thus their estimate of $M_{500} = 5.7^{+3.4}_{-2.1} \times 10^{14} M_{\odot}$ are also consistent with our results (with an $r_{500}$ radius of 1070 kpc compared to 1089 kpc here)

\section{Conclusion}

In this paper, we use Chandra X-ray data, MUSE IFU spectroscopy, and HST imaging to model the overall mass and galactic substructure in the strong lensing cluster MACS J0138.0-2155, which hosts two strongly lensed supernova. We find the central redshift of the cluster to be $z=0.3367$ and find a slightly higher redshift, small group of five galaxies at $z=0.37$ which includes the massive galaxy A.  Galaxy A lies close in projection to the cluster central galaxy and has a stellar velocity dispersion of $291 \pm 3$ km s$^{-1}$.

The X-ray data show MASCJ0138 to be round and relaxed with an ellipticity within the lensing region of $e=0.12\pm0.03$ and a small core radius potentially indicating that this is a cool-core cluster.  The X-ray temperature $6.7\pm0.4$ keV indicates a cluster mass within $r_{500}$ of $5.2^{+1.5}_{-1.2} \times 10^{14} M_{\odot}$.

For the overall cluster velocity dispersion, we find a bootstrap median and spread of $718^{+132}_{-182}$ km s$^{-1}$ for the gapper method consistent with the biweight estimate and result for the nominal sample.  This velocity dispersion gives an estimated mass of $b_\nu^{1/\alpha} M_{200} = 3.6^{+2.0}_{-2.7} \times 10^{14} M_\odot$, subject to the unknown bias in the galactic orbits $b_v$.  This mass is somewhat lower than, but consistent with the X-ray estimated mass.

Using 18 of the brightest, non-blended and quiescent cluster galaxies, we also model the relationship between the luminosity and stellar velocity dispersion of individual member galaxies.  This cluster specific Faber-Jackson relation allows us to benchmark the mass contributions of the satellite galaxy population within the cluster. We run a series of systematics tests to estimate the effect of our analysis choices in modeling the stellar dispersions on the results. We find that the systematics errors are less than the statistical ones, but still significant.  For the slope of the Faber-Jackson relation we find $\alpha = 0.26 \pm 0.06 (\mathrm{stat.}) \pm 0.03 (\mathrm{sys.})$ with an intrinsic scatter of $\Delta \sigma = 31^{+8}_{-6} (\mathrm{stat.}) \pm 4 (\mathrm{sys.})$ km s$^{-1}$ at a reference velocity dispersion of $223$ km s$^{-1}$.

The findings of this work are largely in agreement with similar independent works, such as \citet{Acebron:2025} and \citet{Granata24}.

\section{Acknowledgments}

Parts of this work were funded by NASA grant number GO3-24103X. Some of the data presented in this paper were data obtained from the ESO Science Archive Facility and other data were obtained from the Mikulski Archive for Space Telescopes (MAST). STScI is operated by the Association of Universities for Research in Astronomy, Inc., under NASA contract NAS5-26555. Support for MAST for non-HST data is provided by the NASA Office of Space Science via grant NNX13AC07G and by other grants and contracts. This research has also made use of data obtained from the Chandra Data Archive provided by the Chandra X-ray Center (CXC).

\section{Appendix} \label{sec:Appendix}

In this Appendix we show histograms of the best-fit parameters for the Faber-Jackson relation from our 81 systematics runs, described in Section \ref{sec:FJmethod}.

\begin{figure}
    \centering
    \includegraphics[width=1\linewidth]{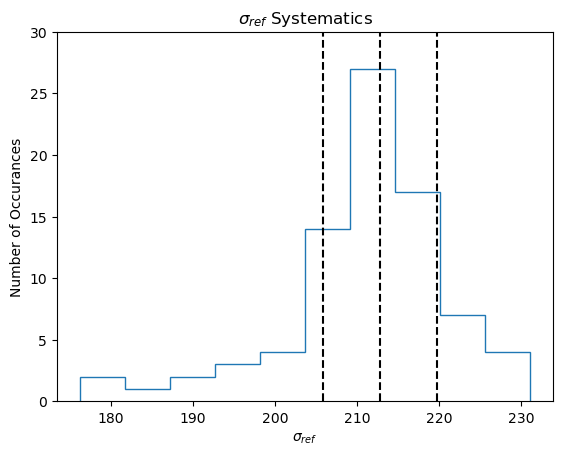}
    \caption{Histogram showing the distribution of best-fit values for $\sigma_{\mathrm{ref}}$, the reference velocity dispersion in the Faber-Jackson relation, derived from the systematics tests.}
    \label{fig:Sigma_Systematics}
\end{figure}

\begin{figure}
    \centering
    \includegraphics[width=1\linewidth]{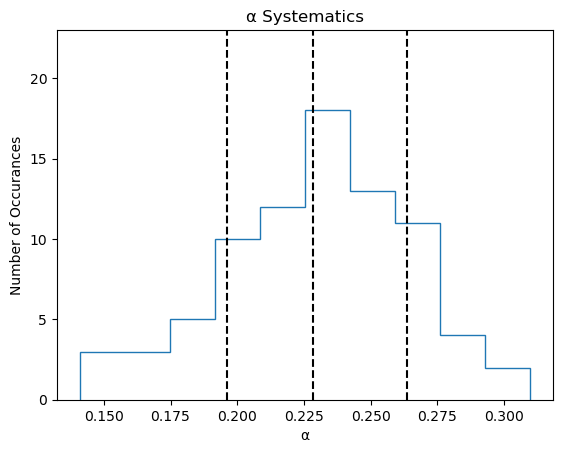}
    \caption{Histogram showing the distribution of best-fit values for $\alpha$, the slope of the Faber-Jackson relation, derived from the systematics tests.}
    \label{fig:Alpha_Systematics}
\end{figure}

\begin{figure}
    \centering
    \includegraphics[width=1\linewidth]{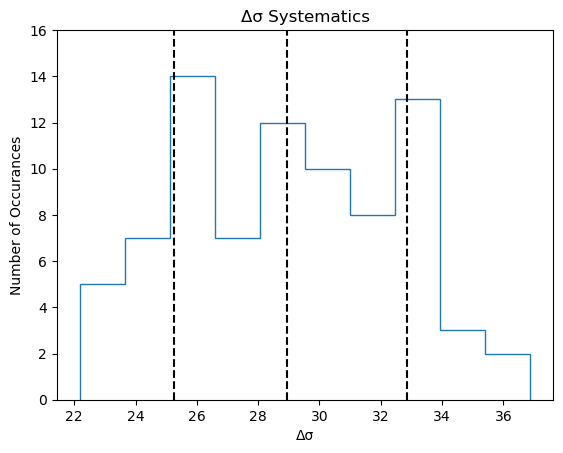}
    \caption{Histogram showing the distribution of best-fit values for $\Delta\sigma$, the intrinsic scatter in the Faber-Jackson relation, derived from the systematics tests.}
    \label{fig:DSigma_Systematics}
\end{figure}

\pagebreak

\bibliography{sources}{}
\bibliographystyle{aasjournal}

\end{document}